\documentclass[12pt]{article}

\usepackage{graphics,amssymb,epsfig,float}
\usepackage[usenames,dvips]{color}
\usepackage{graphicx}
\usepackage{epsfig}
\usepackage{rotating}
\usepackage{dcolumn}
\usepackage{bm}
\usepackage{cite}
\usepackage{amsmath}
\usepackage{array}
\usepackage{setspace}

\makeatletter

\@addtoreset{equation}{section}
\makeatother

\def\lsim{\;\raise0.3ex\hbox{$<$\kern-0.75em\raise-1.1ex\hbox{$\sim$}}\;}
\def\gsim{\;\raise0.3ex\hbox{$>$\kern-0.75em\raise-1.1ex\hbox{$\sim$}}\;}
\def\beq{\begin{equation}}   \def\eeq{\end{equation}}
\def\ba{\begin{array}}       \def\ea{\end{array}}
\def\bea{\begin{eqnarray}}   \def\eea{\end{eqnarray}}

\def\nl{\newline}

\begin{document}

\begin{titlepage}
\begin{flushright}
LPT Orsay 14-91 \\
PCCF RI 14-09\\
\end{flushright}


\begin{center}

\begin{doublespace}

\vspace{1cm}
{\Large\bf Excessive Higgs pair production with little MET from squarks
and gluinos in the NMSSM} \\
\vspace{2cm}

{\bf{Ulrich Ellwanger$^a$ and Ana M. Teixeira$^b$}}\\
\vspace{1cm}
{\it  $^a$ LPT, UMR 8627, CNRS, Universit\'e de Paris--Sud, 91405 Orsay,
France, and \\
\it School of Physics and Astronomy, University of Southampton,\\
\it Highfield, Southampton SO17 1BJ, UK\\
\it $^b$ Laboratoire de Physique Corpusculaire, CNRS/IN2P3 - UMR 6533\\
Campus des C\'ezeaux, 24 Av. des Landais, F-63171 Aubi\`ere, France }
\end{doublespace}
\end{center}
\vspace*{2cm}

\begin{abstract}
In the presence of a light singlino-like LSP in the NMSSM, 
the missing transverse energy -- MET -- 
signature of squark/gluino production can be considerably reduced.
Instead, a pair of Higgs bosons is produced in each event. 
We propose benchmark
points for such scenarios, which differ in the squark and gluino masses,
and in their decay cascades. Events for these points are simulated for the
run~II of the LHC at 13~TeV centre of mass energy. 
After cuts on the transverse momenta of at
least four jets, and requiring two $\tau$-leptons from one Higgs decay,
we find that the invariant mass of two $b$-jets from the other Higgs
decay shows clear peaks above the background. Despite the reduced MET,
this search strategy allows to see signals for sufficiently large 
integrated luminosities, depending on the squark/gluino masses.
\end{abstract}

\end{titlepage}

\section{Introduction}
After the first run of the LHC at a centre of mass (c.m.) energy of mostly
8~TeV, no significant excesses have been observed in searches for physics
beyond the Standard Model (SM). Amongst others, this concerns searches for
supersymmetric particles (sparticles) like squarks, gluinos, electroweak
gauginos and higgsinos. 

Lower bounds on the masses of sparticles have been 
obtained~\cite{atlas_summary,cms_summary} 
which depend, however, on the sparticle
decay cascades and hence on the complete sparticle spectrum. Recent summaries of
bounds within various scenarios can be found 
in~\cite{Feng:2013pwa,Craig:2013cxa,Melzer-Pellmann:2014eta,Halkiadakis:2014qda}. 
In particular, for similar squark\footnote{Subsequently we use the notion
``squark'' for the scalar partners
of the quarks of the first two generations; the scalar partners of top
and bottom quarks will be denoted by stops and sbottoms, respectively.}
masses $M_{\tilde{q}}$ and gluino masses  $M_{\tilde{g}}$ and decay
cascades motivated by the Minimal Supersymmetric extension of the
Standard Model (MSSM), ATLAS obtained
$M_{\tilde{q}} \sim M_{\tilde{g}} \gsim 1.7$~TeV~\cite{Aad:2014wea}.
Weaker limits are obtained within simplified models where, for instance,
gluinos are assumed to be decoupled in the case of squark production.
(Decoupled gluinos imply reduced squark production cross sections and,
for similar squark and gluino masses,
the largest production cross sections are the ones corresponding
to one squark plus one gluino production.) In any case, these lower
bounds have already put the MSSM under a certain stress.

Searches for squarks, gluinos and sparticles rely in general (assuming
conserved R-parity) on events with large missing transverse energy
$E_T^\mathrm{miss}$ (MET) due to the escaping stable lightest supersymmetric
particle (LSP), which is a good dark matter candidate if neutral. Since
corresponding cuts on $E_T^\mathrm{miss}$ are applied, these searches are
less effective if, for kinematical reasons, the LSP produced in the
last step $\text{NLSP} \to X+\text{LSP}$ of a sparticle decay chain is
always soft and carries little energy. (NLSP denotes the Next-to-lightest
supersymmetric particle.)

This is the case if the LSP is very light ($M_\mathrm{LSP} \sim$ few GeV)
and, simultaneously, $X$ corresponds to a Standard Model particle, with
$M_X$ close to (just below) $M_\mathrm{NLSP} - M_\mathrm{LSP}$. A possible
Standard Model particle $X$ is the 125~GeV Higgs boson, although additional
non-SM-like Higgs bosons (or the $Z$ boson) could also play
that role.
In the general MSSM such a scenario cannot be realised in practise:
Whereas a bino-like LSP could be in principle very light and a wino- or
higgsino-like NLSP have a mass close to 125 GeV + $M_\mathrm{LSP}$,
most squarks (and sleptons) would then prefer to decay directly into the
LSP, skipping the step $\text{NLSP} \to X+\text{LSP}$. However, if this step
is not present in nearly all sparticle decays, existing lower bounds on
sparticle masses are hardly alleviated due to the other decay processes
with an energetic LSP and large $E_T^\mathrm{miss}$.

On the other hand, in extensions of the MSSM it is possible that the LSP
has only weak couplings to all sparticles present in the MSSM. Then the 
MSSM-like sparticles (squarks, gluinos etc.) avoid direct decays into the
LSP, but their decay cascades end up (provisionally) in the
``MSSM-like LSP'', typically the bino. Only subsequently does 
the ``MSSM-like LSP'' (now the NLSP) decay into the ``true''
LSP~$+ X$, always leading to a soft LSP for configurations of masses as
stated above.

Scenarios of that kind have been discussed in~\cite{Fan:2011yu,Lisanti:2011tm,
Fan:2012jf, Baryakhtar:2012rz,Evans:2013jna,Ellwanger:2014hia}.
The role of the ``true'' LSP can be
played by a light gravitino (provided the decay of the MSSM-like LSP
happens inside the detector -- otherwise it behaves like the true 
LSP)~\cite{Fan:2011yu,Fan:2012jf}, so-called photini~\cite{Baryakhtar:2012rz},
or the singlino of the NMSSM~\cite{Ellwanger:2014hia}. 

The NMSSM denotes the Next-to-Minimal Supersymmetric extension of the Standard
Model~\cite{Ellwanger:2009dp} where the coupling of the two Higgs
doublets of the MSSM to an additional gauge singlet field $S$ solves
the $\mu$-problem of the MSSM~\cite{Kim:1983dt}, and renders more
natural a value of $\sim 125$~GeV of the SM-like Higgs 
boson~\cite{Hall:2011aa,Ellwanger:2011aa,Arvanitaki:2011ck,King:2012is,
Kang:2012sy,Cao:2012fz}, while preserving the attractive features of the
MSSM. Besides the Higgs sector, the NMSSM differs from the MSSM through
the presence of an additional neutralino (the singlino, the fermionic
component of the singlet superfield). The singlino can be a light LSP,
weakly coupled to the MSSM-like sparticles. A scenario with such a soft
singlino-like LSP in the NMSSM was also briefly discussed 
in~\cite{Lisanti:2011tm} and in a variant of the
NMSSM including non-renormalisable terms in~\cite{Fan:2011yu}.

In~\cite{Ellwanger:2014hia} we studied in detail to which extent the
reduction of $E_T^\mathrm{miss}$ due to a light singlino in the NMSSM can
alleviate the lower bounds on squark and gluino masses from the run~I of
the LHC. We presented a ``worst case scenario'' with all sparticle masses
below $\sim 1$~TeV, but consistent with constraints from the LHC. The
dominant limits on such scenarios with little $E_T^\mathrm{miss}$
actually originate from searches for many hard jets as 
in~\cite{ATLAS-CONF-2013-091,CMS-PAS-EXO-12-049,Chatrchyan:2013izb,
Chatrchyan:2013xva}. One particular feature of the scenario presented
in~\cite{Ellwanger:2014hia} is that the role of $X$ is played by a
NMSSM-specific Higgs boson lighter than $M_Z$ (but not ruled out by LEP).
Such light Higgs bosons have very small branching ratios ($BR$s)
into $W^{(*)}W^{(*)}/Z^{(*)}Z^{(*)}$ with subsequent leptonic decays of
$W^{(*)},\ Z^{(*)}$
leading to neutrinos. Neutrino decays generate $E_T^\mathrm{miss}$,
which makes corresponding scenarios somewhat more sensitive to standard
supersymmetry search channels.

The production of Higgs bosons from neutralino cascades has been studied
before in variants of the MSSM~\cite{Dimopoulos:1996vz,Ambrosanio:1996jn,
Hinchliffe:1996iu,Matchev:1999ft,Datta:2003iz,Bandyopadhyay:2008fp,
Huitu:2008sa,Bandyopadhyay:2008sd,Fowler:2009ay,Meade:2009qv,Asano:2010ut,
Thaler:2011me,Gori:2011hj,Ruderman:2011vv,Kats:2011qh,Baer:2012ts,
Ghosh:2012mc,Belyaev:2012si,Byakti:2012qk,Baryakhtar:2012rz,Howe:2012xe,
Belyaev:2012jz,Arbey:2012fa,Bharucha:2013epa,Han:2013kza,Yu:2014mda}
and the NMSSM~\cite{Franke:1995tf,Ellwanger:1997jj,Choi:2004zx,
Cheung:2008rh,Stal:2011cz,Das:2012rr,Cerdeno:2013qta}. The possible
reduction of $E_T^\mathrm{miss}$ due to a softer LSP in such decays was
observed in~\cite{Ruderman:2011vv,Kats:2011qh,Howe:2012xe,Das:2012rr},
but the emphasis was mainly on neutralino decays as additional sources
of Higgs bosons. Since these can be considerably boosted, analyses of
the substructure of the resulting jets have been proposed 
in~\cite{Kribs:2009yh,Kribs:2010hp,Byakti:2012qk,Bhattacherjee:2012bu}.

In the present paper we concentrate, in contrast to~\cite{Ellwanger:2014hia}, 
on the possible excessive production of pairs
of Standard Model-like 125~GeV Higgs bosons, $H_{SM}$. The ``worst case
scenarios'' discussed in~\cite{Ellwanger:2014hia} relied, for reasons
stated above, on the production of a lighter NMSSM-specific Higgs boson
$H_1$ in the $\text{bino} \to X+\text{singlino}$ cascade and, moreover,
it was assumed that squarks $\tilde{q}$ directly decay into quarks and
the bino in order to alleviate as much as possible 
the constraints from searches based on
$E_T^\mathrm{miss}$ and jets. Here we study scenarios
with longer squark decay cascades: squarks decaying via gluinos and/or
gluinos decaying via stops/sbottoms. The final step in the decay
cascades is always assumed to be $\text{bino} \to H_{SM}+\text{singlino}$,
hence $E_T^\mathrm{miss}$ is still strongly reduced, making standard
searches for supersymmetry less efficient. The aim is then to see whether
signals of two Standard Model-like 125~GeV Higgs bosons $H_{SM}$ can be
extracted (above the Standard Model background) in order not to miss
squark/gluino production at the 13~TeV c.m. energy run~II at the LHC.

To this end we present benchmark points with squark/gluino masses in the
1-1.5~TeV range, which are not excluded by searches from run~I. The
benchmark points differ in the decay cascades; if stops/sbottoms appear
therein, their masses are assumed to be $\sim 750$~GeV (the precise
values of their masses have little impact on the final signatures).
In each case we perform simulations and attempt to extract signals
of two Standard Model-like 125~GeV Higgs bosons $H_{SM}$ above the
background, concentrating as in~\cite{Ellwanger:2014hia} on final states
with $2\ \tau's$ and a $b\bar{b}$ pair, the invariant masses of the latter
near the Higgs mass.

In the next section we present the scenarios and the corresponding
benchmark points in more detail. In Section~3 we describe the simulations,
the analysis and the dominant backgrounds. In Section~4 we collect the
results for the benchmark points and discuss which of their properties
can help to distinguish the various scenarios. Section~5 is devoted to
a summary and conclusions.

\section{Scenarios with little MET in the NMSSM}

As described in the Introduction and discussed in more detail 
in~\cite{Ellwanger:2014hia}, a loss of $E_T^\mathrm{miss}$ (associated
to the LSP)
in sparticle decay cascades can occur in the NMSSM if a dominantly
singlino-like LSP is light, the mass of the typically
mostly bino-like NLSP $M_\mathrm{NLSP}$ 
is somewhat above the sum of a Higgs and LSP masses 
$M_\mathrm{NLSP} \gtrsim M_H + M_\mathrm{LSP}$, 
and practically all sparticle decay cascades terminate
by a last step $\text{NLSP} \to H + \text{LSP}$. (Decays
$\text{NLSP} \to Z + \text{LSP}$ occur only if the NLSP has a higgsino
component, i.e. if the effective $\mu$ parameter $\mu_\text{eff}$ of the
NMSSM~\cite{Ellwanger:2009dp} is relatively small, which we do not
assume here for simplicity.)

The case where $H$ corresponds to a NMSSM-specific Higgs scalar with a
mass below $M_Z$ was investigated in~\cite{Ellwanger:2014hia} (in order
to reduce as much as possible $E_T^\mathrm{miss}$ in all decay processes);
here we consider the Standard Model-like $H_{125}$ and, accordingly,
a bino-like NLSP mass somewhat above 125~GeV, as such scenarios are equally
possible in the general NMSSM. 
The main purpose of the present paper is to
propose and study benchmark points for squark/gluino production which,
due to the reduction of $E_T^\mathrm{miss}$, would be difficult to
observe in standard supersymmetry search channels relying on large cuts
on $E_T^\mathrm{miss}$. Instead, we show that -- for not too heavy
squarks and large enough integrated luminosity -- such scenarios are
observable in searches for two Higgs bosons accompanied by hard jets.

The benchmark points considered here include scenarios where squarks
decay via glu\-inos, leading to more jets in the final state but with
reduced transverse momenta of the (s)particles involved in the last
decay step. Gluinos lighter than squarks can undergo 3-body decays into
two quarks and a bino or, if kinematically allowed, 2-body decays into
top-stop or bottom-sbottom pairs. Decay chains involving charginos or
heavier neutralinos are left aside here as their decays via $W^\pm$ or
$Z$ bosons can lead to neutrinos, and the $E_T^\mathrm{miss}$ from 
the latter $\nu$'s is often sufficient to make 
standard supersymmetry searches
(possibly including isolated leptons) sensitive to these scenarios.
Direct production of charginos or heavier neutralinos with subsequent
Higgs pair production instead of $E_T^\mathrm{miss}$ merits a separate
analysis.

The subsequent benchmark points will not be defined in terms of
parameters of the NMSSM but, for convenience, in terms of masses and
branching fractions of the involved sparticles. However, these are chosen
such that they can be reproduced at least approximatively by suitable
parameters of the NMSSM\footnote{This also ensures that 
all phenomenological constraints (except for the muon anomalous 
magnetic moment) tested in NMSSMTools are satisfied, in particular
those from flavour physics (B meson decays).} as we have checked using 
{\sf NMSSMTools\_4.4.0}~\cite{Ellwanger:2004xm,Ellwanger:2005dv}.

In~\cite{Ellwanger:2014hia} it was found that
the loss of $E_T^\mathrm{miss}$ is not very sensitive to the masses of
the particles involved in the decay $\text{NLSP} \to H + \text{LSP}$ as
long as $M_\mathrm{NLSP}-(M_H+M_\mathrm{LSP}) \ll M_\mathrm{NLSP}$,
and we verified that this also holds for the signals obtained below.
Hence we do not vary these masses and choose for all benchmark points
$M_\mathrm{NLSP} = 130$~GeV, $M_H = 125$~GeV, $M_\mathrm{LSP} = 3$~GeV
and 100\%~$BR$ for the decay
$\text{NLSP} \to H_{125} + \text{LSP}$.

For the squark masses $M_{\tilde q}$ and gluino masses $M_{\tilde g}$ we
choose values
such that the benchmark points are not ruled out by searches at the
run~I at the LHC (using CheckMATE~\cite{Drees:2013wra} for the standard
supersymmetry search channels). All squarks of the first two generations
are assumed degenerated. As stated above, the points differ by
their squark~$\to$~LSP decay chains, and for completeness we start with
points for which this chain is as short as possible: decoupled stops
and sbottoms, and gluinos only slightly heavier than -- almost
degenerate with -- squarks. (Squarks much lighter than the gluino are
unstable under radiative corrections and would imply an unnatural tuning
of bare squark mass parameters.)
The squark, gluino and stop/sbottom masses of the eight points P1 -- P8
are shown in Table~1, where we also include the sums of
squark-squark, squark-antisquark, squark-gluino and gluino-gluino cross
sections $\sigma_\text{tot}$ as obtained from Prospino at 
NLO~\cite{Beenakker:1996ch,Beenakker:1996ed}. 

Points P1 and P2 are examples of short decay chains;
for P1 the squark/gluino masses are
slightly above the lower bounds from the LHC run~I (i.e. $\sim 1$~TeV,
taken $\sim 150$~GeV heavier than in the 
``worst case scenario'' studied in~\cite{Ellwanger:2014hia}) 
while the squark/gluino
masses have somewhat more pessimistic values of $\sim 1.4$~TeV
for P2. Both squark and gluino production contribute to the total
production cross section.
Gluinos heavier than squarks are assumed to decay democratically into
all squark-quark pairs of the first two generations, and squarks lighter
than gluinos with 100\% $BR$ into the bino-like NLSP and the
corresponding quark: $\tilde g \to q \, \tilde q \to q \, \chi^0_2$. 

\begin{table}[h!]
\begin{center}
\begin{tabular}{|c|c|c|c|c|}
\hline
Point & $M_{\tilde q}$ [GeV]& $M_{\tilde g}$ [GeV]& $M_{\tilde t}$ or
$M_{\tilde b}$ [GeV] 
& $\sigma_\text{tot}$ [fb] \\
\hline
P1 & 1000 & 1010 & decoupled & $\sim 1645$   \\
\hline
P2 & 1400 & 1410 & decoupled & $\sim 168$ \\
\hline
P3 & 1100 & 900 & decoupled & $\sim 1874$ \\
\hline
P4 & 1500 & 1300 & decoupled & $\sim 169$  \\
\hline
P5 & 1400 & 1410 & $M_{\tilde t}$: 750 & $\sim 168$ \\
\hline
P6 & 1100 & 1110 & $M_{\tilde b}$: 750 & $\sim 920$ \\
\hline
P7 & 1500 & 1300 & $M_{\tilde t}$: 750 & $\sim 169$ \\
\hline
P8 & 1400 & 1200 & $M_{\tilde b}$: 750 & $\sim 321$ \\
\hline
\end{tabular}
\caption{Squark, gluino, stop and sbottom masses (unless decoupled) and
the sum of squark-gluino production cross sections at NLO for the
benchmark points P1 -- P8.}
\end{center}
\end{table}

Points P3 and P4 correspond to scenarios where gluinos are lighter
than squarks. Now one can assume that the left-handed squarks
fully decay into gluinos and the corresponding quarks 
$BR (\tilde q_L \to q \tilde g )\sim$ 100\%, 
while right-handed squarks, due to the
larger hypercharge, have $BR$ of about 30\% (assumed to be precise)
into the bino-like NLSP and the corresponding quarks, leaving 70\%~$BR$
into gluinos and the corresponding quarks, 
$BR (\tilde q_R \to q \chi^0_2 )\sim$ 30\%, 
$BR (\tilde q_R \to q \tilde g)\sim$ 70\%.

For the remaining points P5--P8 
we assume stops or sbottoms lighter than gluinos, in fact 
lighter than $M_{\tilde g} - m_\text{top}$ to allow for gluino
2-body decays. As before, gluinos can be slightly heavier or lighter
than quarks, but squark and gluino masses have to be somewhat larger 
(depending on whether we have light stops or sbottoms) in
order to be compatible with the limits from the run~I of the LHC. 
For the same reason, 
stop or sbottom masses should be large enough. We observed that,
as long as gluinos decay with 100\% $BR$ into top + stop or bottom + sbottom,
the signals depend very little on the stop/sbottom masses, 
provided these are below
$M_{\tilde g} - m_\text{top}$. Hence, instead of varying the stop/sbottom
masses, we choose a sufficiently large value, 
$M_{\tilde t, \tilde b}=750$~GeV to comply with current LHC
limits, but we neglect contributions from stop/sbottom pair
production to the signals. These contributions are found to be very
small (after the cuts discussed below) and would decrease even further
for heavier stops/sbottoms; hence the signal rates also remain valid for
heavier stops/sbottoms.

Points P5 and P6 correspond again to gluinos slightly heavier than
squarks: 
for the gluino, $BR(\tilde g \to t \tilde t)\sim 100\%$ (P5) or 
$BR(\tilde g \to b \tilde b)\sim 100\%$ (P6), 
and squarks decay with 100\% $BR$
into the bino-like NLSP and the corresponding quark, 
$BR(\tilde q \to q \chi^0_2) \sim 100\%$.

For the points P7 and P8, gluinos are lighter than squarks. Again,
right-handed squarks are assumed to decay partially (with 70\%~$BR$)
into gluinos and the corresponding quark, with a $BR$ of 30\% into the
bino-like NLSP and the corresponding quark, but left-handed squarks with
$BR \sim 100\%$ into gluinos. Herewith all relevant
masses (summarised in Table~1) and branching fractions of the benchmark
points are defined.

\section{Extraction of signals in $b\bar{b}+\tau^+\tau^-$ final states}

Events due to squark/gluino production from $pp$ collisions at 13~TeV are
simulated using MadGraph/\-MadEvent~\cite{Alwall:2011uj} which
includes Pythia~6.4~\cite{Sjostrand:2006za} for showering and
hadronisation. The emission of one additional hard jet was allowed in the
simulation; the production cross sections for the four distinct
squark-squark, squark-gluino, squark-antisquark and gluino-gluino
production processes were obtained separately by Prospino at 
NLO~\cite{Beenakker:1996ch,Beenakker:1996ed}. (The dominant contributions
always come from squark-squark and squark-gluino production.)
The output was given in StdHEP format to the detector simulation 
DELPHES~\cite{deFavereau:2013fsa}. Jets were constructed by 
Fastjet~\cite{Cacciari:2011ma} (part of the Delphes package) using the
anti-$k_T$ algorithm~\cite{Cacciari:2008gp}. For $b$-tagged jets we
require $p_T > 40$~GeV and assume a $b$-tag efficiency of 70\%
(mistag efficiencies from $c$-jets of 10\%, and from light quark/gluon
jets of 1\%).

For the analysis we try to profit from the fact that the events are
rich in hard jets, and each event contains two Higgs bosons. A strong
reduction of the Standard Model background -- keeping the signal
acceptance as large as possible -- is achieved (see below) if
we require that signal events contain at least two hadronically decaying
$\tau$ leptons and at least two $b$-tagged jets.

\begin{figure}[ht!]
\begin{center}
\hspace*{-6mm}
\begin{tabular}{cc}
\psfig{file=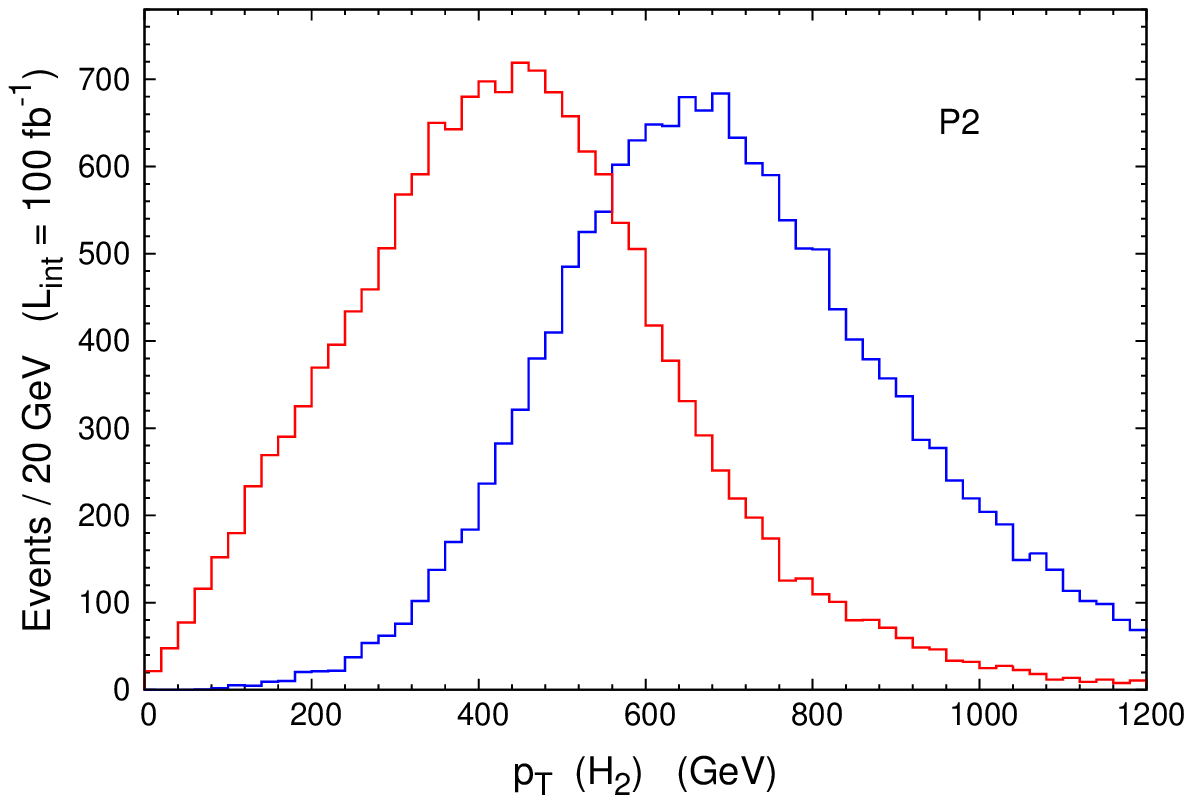, scale=0.60}
\   &
\psfig{file=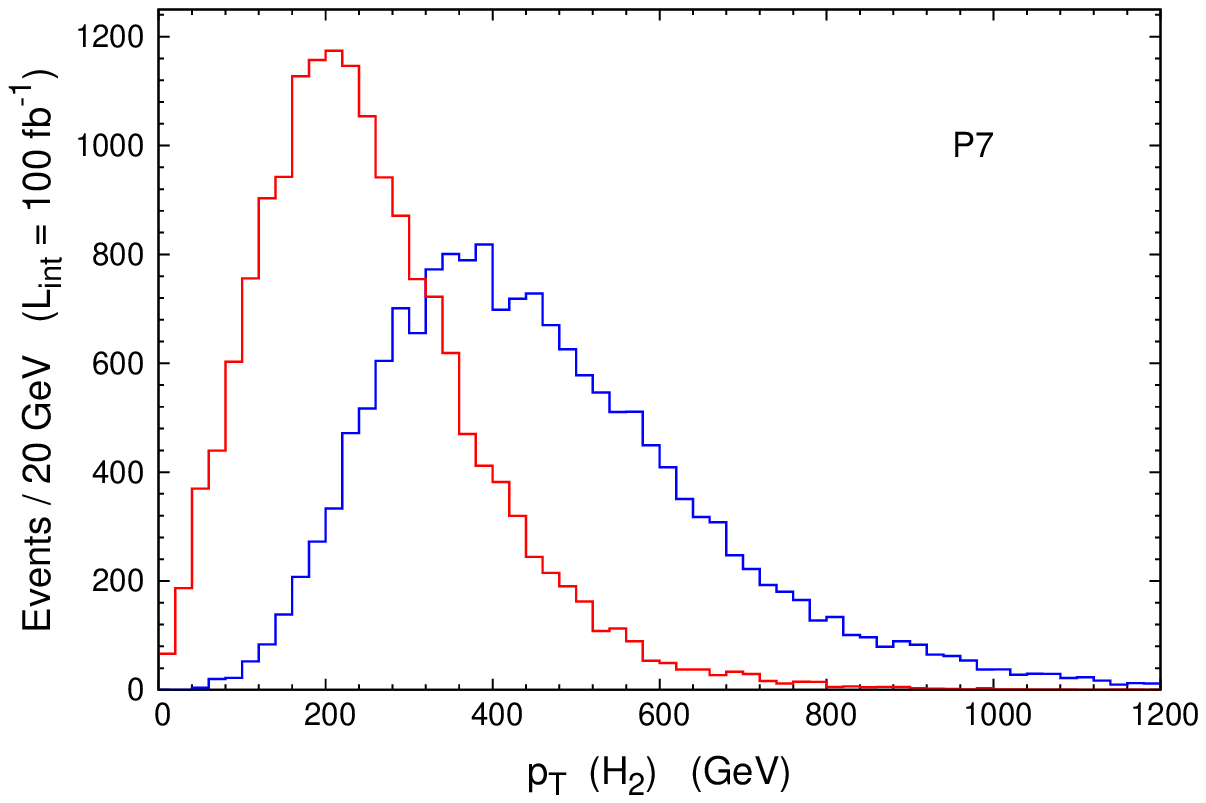, scale=0.60}
\end{tabular}
\end{center}
\caption{Spectra of the transverse momenta of the leading (blue) and
next-to-leading (red) Higgs bosons for the benchmark points P2 (left panel)
and P7 (right panel).}
\label{fig:1}
\end{figure}

The average transverse momentum of the Higgs bosons depends on the
squark/gluino masses and, notably, on the length of the decay chains.
This is clarified in comparing the spectra of the transverse momenta
of the leading and subleading Higgs bosons of the points P2 and P7 in
Fig.~\ref{fig:1}: The masses of the originally produced squarks and
gluinos are similar ($\sim 1.4$~TeV for P2, 1.3/1.5~TeV for P7), but the
transverse momenta of the leading Higgs boson peak near 700~GeV for P2,
where squarks decay directly into the NLSP, but somewhat below 400~GeV
for P7, in which one finds longer cascades, with the 
squarks decaying into gluinos and stops. (These analyses were
performed by means of MadAnalysis~5~\cite{Conte:2012fm,Conte:2013mea}.)

In principle, the decay products of strongly boosted Higgs bosons, as is
the case of point P2, can be analysed using jet substructure 
methods~\cite{Kribs:2009yh,Kribs:2010hp,Byakti:2012qk,Bhattacherjee:2012bu}.
A similar approach, based on the construction of ``slim'' $R=0.15$ jets
for each event, was employed in~\cite{Ellwanger:2014hia} where only
scenarios with short squark decay cascades were considered. However, 
we found that the latter approach fails for scenarios with long squark
decay cascades, which typically lead to less boosted Higgses.  
Instead, a more standard method leads to satisfactory results:

We construct jets with a jet cone radius $R=0.4$, and require at least
two such jets to be $b$-tagged. Then we define the invariant mass $M_{bb}$
of the two $b$-tagged jets which are closest in $\Delta R$. (Muons inside
such jets are added to the invariant mass of the system.) This simple
approach works for all benchmark points; even the ones with large
average transverse momenta of Higgs bosons lead to sufficiently many
events with less boosted Higgses whose mass can be reconstructed this way.

Compared to analyses based on slim jets, the use of more standard $R=0.4$
jets has the additional advantage that the $\tau$ fake rate (and notably
the unusually large $2$-$\tau$ fake rate observed in~\cite{Ellwanger:2014hia})
is much smaller, which helps to suppress the background from QCD.

\begin{figure}[ht!]
\begin{center}
\hspace*{-6mm}
\begin{tabular}{cc}
\psfig{file=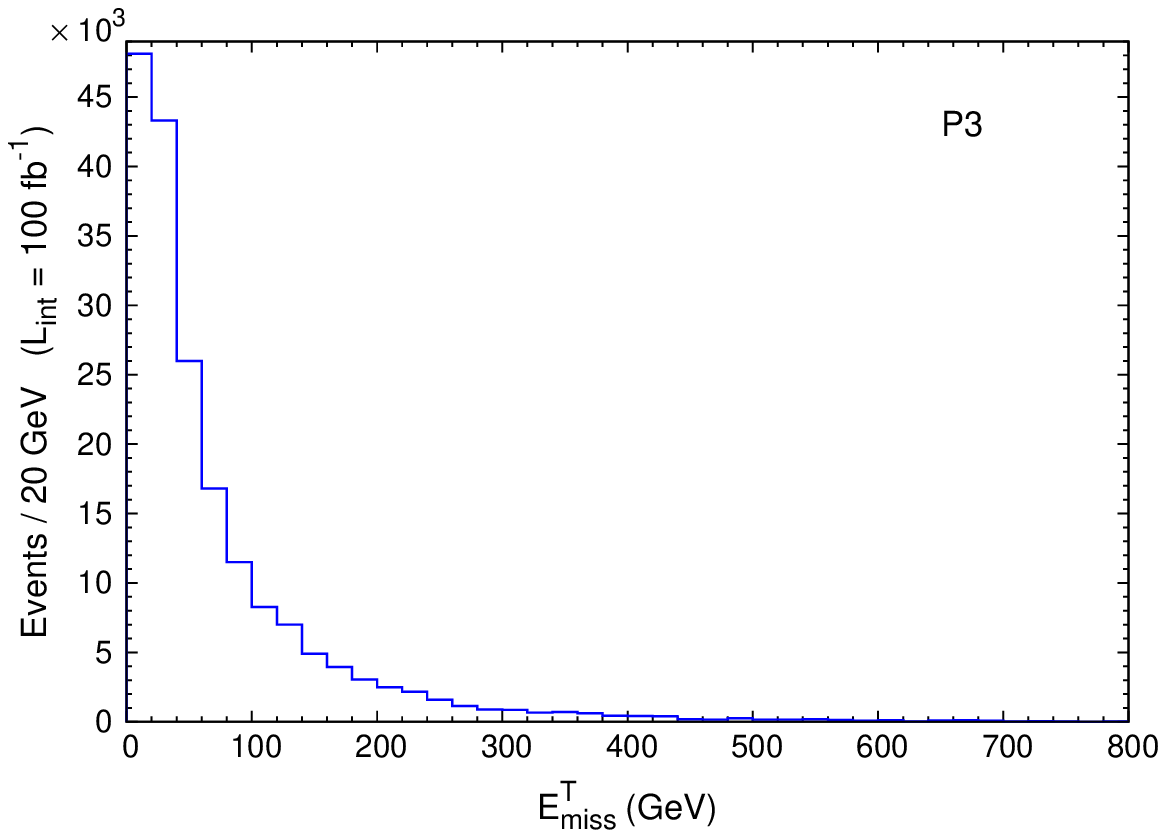, scale=0.60}
\   &
\psfig{file=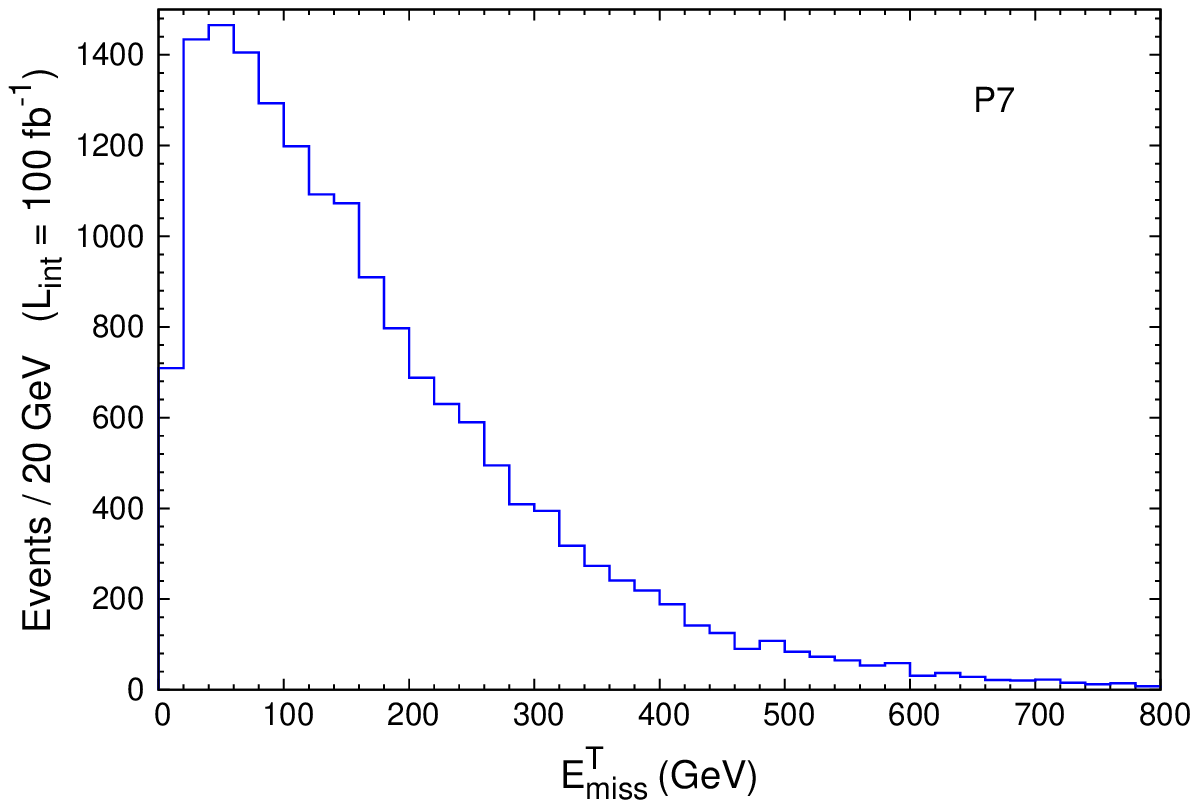, scale=0.60}
\end{tabular}
\end{center}
\caption{Spectra of $E_T^\mathrm{miss}$ for the benchmark points P3
(left panel) and P7 (right panel).}
\label{fig:2}
\end{figure}

The spectra of $E_T^\mathrm{miss}$ are quite soft for all 
the benchmark points due to the kinematical reasons discussed in 
the Introduction. Still, requiring two $\tau$'s
in the final state implies that 
some $E_T^\mathrm{miss}$ will always be due to the
escaping $\tau$ neutrinos. In addition, leptonic $b$-decays (from Higgs
decays or cascades via stops/sbottoms) can generate some
$E_T^\mathrm{miss}$. Especially for cascades via top quarks, as in 
points P5 and P7, $E_T^\mathrm{miss}$ from leptonic top decays can be
relatively large. Finally, also for short decay cascades, the
$E_T^\mathrm{miss}$ spectrum becomes harder for heavier squarks and gluinos.
Among the benchmark points, the softest $E_T^\mathrm{miss}$ spectrum
is observed for the point P3, the hardest for the point P7. Both are
shown in Fig.~\ref{fig:2}. The point P7 could possibly also be
discovered in standard search channels with sizeable cuts on
$E_T^\mathrm{miss}$, $E_T^\mathrm{miss} \gsim 100$~GeV, 
but if such cuts were imposed, then 
most of the events from points like P3 would
be missed. Hence we use only a mild lower cut on
$E_T^\mathrm{miss}$ of 30~GeV.

In our analysis the following cuts were applied:
\begin{itemize}
\item 4 jets ($b$-tagged or not) with transverse momenta $P_T$
$> 400$~GeV,
$> 300$~GeV, $> 200$~GeV and $> 100$~GeV, respectively.
\item At least two $b$-jets with $P_T > 40$~GeV were required,
and a small lower cut on $E_T^\mathrm{miss}>30$~GeV was applied.
\item At least two hadronically decaying $\tau$'s are required, with
invariant masses ranging from 20~GeV to 160~GeV, the sum of their transverse
momenta imposed to be above 100~GeV (which further suppresses fake $\tau$'s).
\item Finally a (large) signal region 60~GeV~$<M_{bb}<160$~GeV was
defined; not only this allows 
to take into account uncertainties in the measurements of $M_{bb}$, 
but also to remain sensitive to possible additional
Higgs bosons with masses below 125~GeV. (Additional Higgs bosons with
masses below 60~GeV must be practically pure singlets to avoid
constraints from LEP, and to avoid decays of the 125~GeV Higgs boson
into such pairs which would reduce its observed signal rates. Then the
couplings of such light additional Higgs bosons to a bino must be very
small.)
\end{itemize}

\noindent
Various SM backgrounds have been considered: top quark pair
production, possibly together with 1-2 hard jets at the parton level;
bottom quark pair production, also possibly together with 1-2 hard jets at
the parton level; direct production of $\tau$'s from $Z$ and $W$ bosons
together with QCD jets. The by far dominant contributions to the signal
region were found to originate from top quark pair production together
with 1 hard jet at the parton level (and possibly fake $\tau$'s), and
bottom quark pair production with 2 hard jets at the parton level
(and two fake $\tau$'s).
 We have simulated 300 000 top pair production events and 500 000 bottom pair
production events using the same procedure as for the benchmark points. 
About 0.33\% of the top pair events contained two $\tau$'s satisfying
the criteria of our cuts, while this only occurs for circa 0.03\% of the
bottom pair events. 
Finally we obtained contributions to the
signal region of  $\sim 0.029$~fb from top pair production and
$\sim 0.031$~fb from  bottom pair production, i.e. $\sim 0.06$~fb all
together.

\section{Signals for benchmark points}

In this section we discuss the properties of the benchmark points, the
signals of Higgs bosons as well as other observables which allow for hints
on the underlying sparticle spectrum.

\begin{figure}[b!]
\begin{center}
\hspace*{-6mm}
\begin{tabular}{cc}
\psfig{file=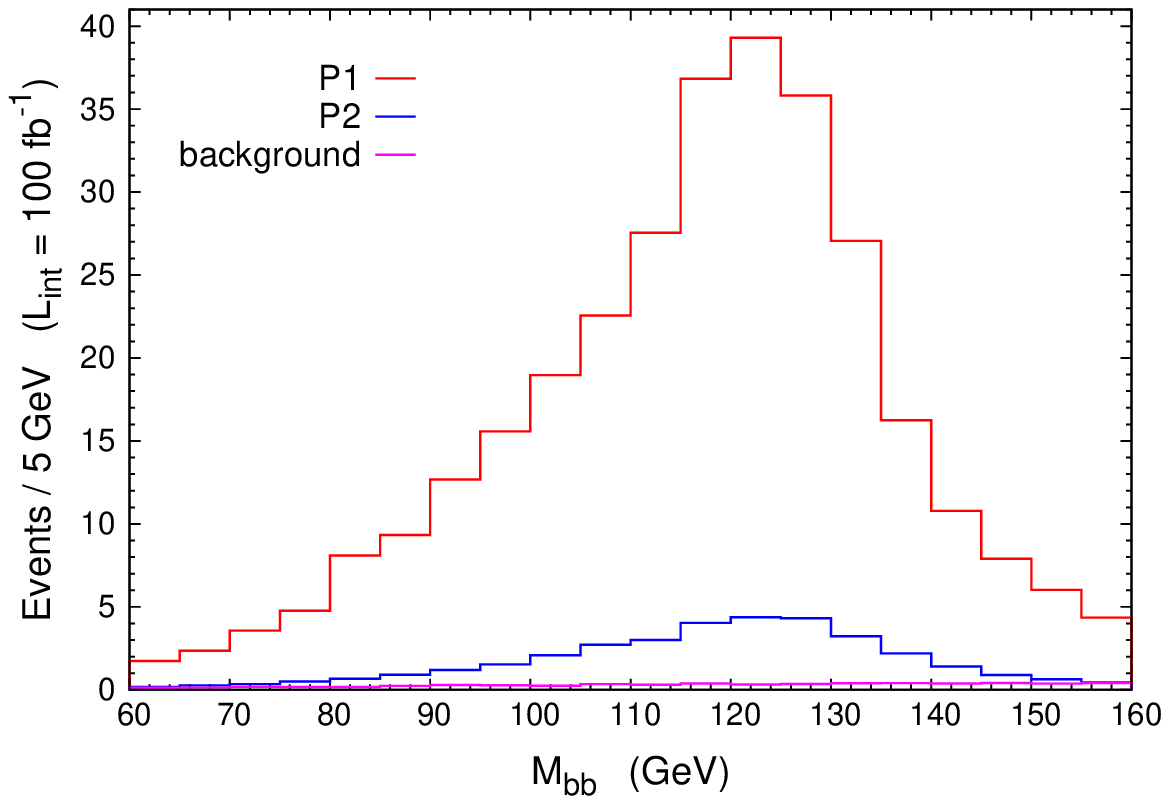, scale=0.60}
\   &
\psfig{file=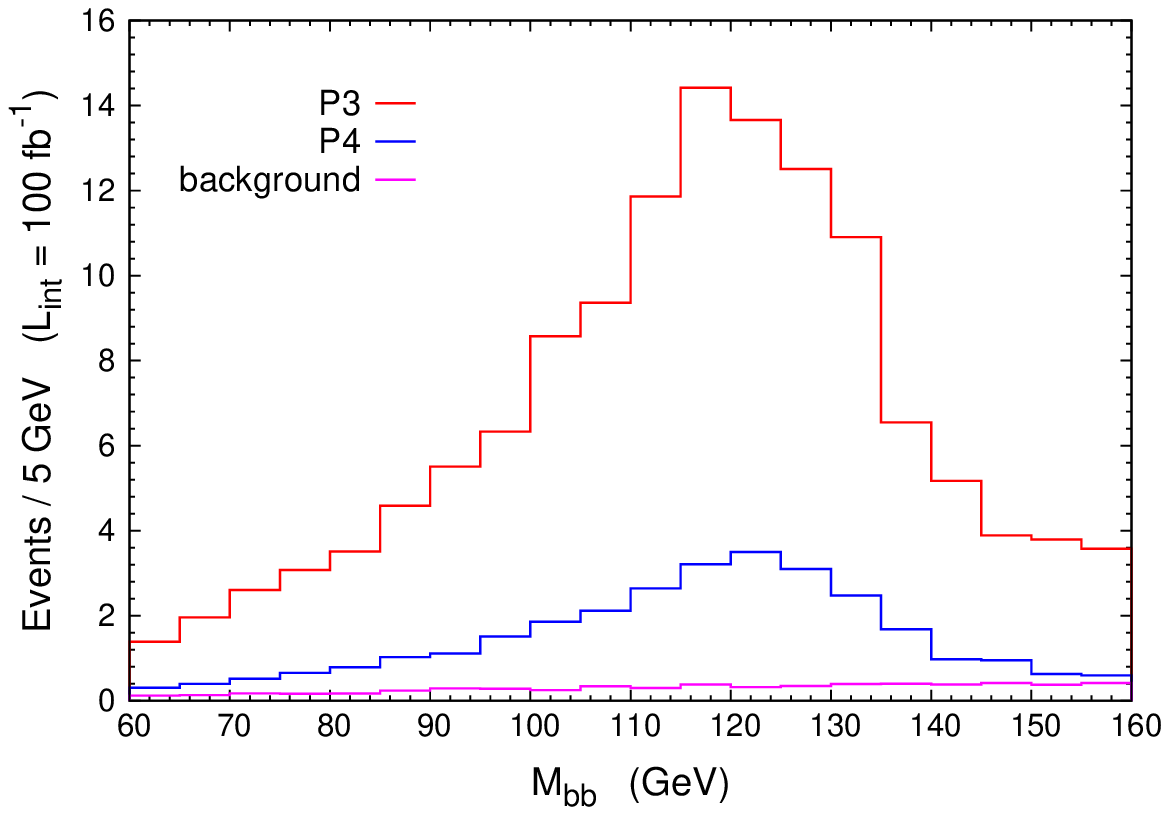, scale=0.60}
\end{tabular}
\end{center}
\caption{Spectra of $M_{bb}$ for the benchmark points P1 and P2
(left panel), and P3 and P4 (right panel).}
\label{fig:3}
\end{figure}

First, the spectra of $M_{bb}$ after the cuts described in Section~3 are
summarised in Fig.~\ref{fig:3} for the points P1 -- P4. The cross
sections in the signal region for these points are $\sim 3.1$~fb~(P1),
$\sim 0.35$~fb~(P2), $\sim 1.3$~fb~(P3) and $\sim 0.30$~fb~(P4), i.e. all
are well above the background cross section in the signal region of
$\sim 0.06$~fb. Comparing the signal cross sections
with the production cross sections in Table~1, we see that the
acceptances are about 1-2$\times 10^{-3}$, increasing with increasing
squark/gluino masses (i.e. more hard jets). This information is
summarised in Table~\ref{table:P:signal}.

\begin{table}[h!]
\begin{center}
\begin{tabular}{|c|c|c|c|c|c|c|c|c|}
\hline
Point & P1 & P2 & P3 & P4 & P5 & P6 & P7 & P8  \\
\hline
$\sigma_\text{signal}$ [fb] & 
3.1 & 0.35 & 1.3 & 0.30 & 0.45 & 2.0 & 0.56 & 0.46 \\
\hline
accept. [$\times 10^{-3}$] & 
1.9 & 2.1 & 0.7 & 1.8 & 2.7 & 2.1 & 3.3 & 1.4  \\
\hline
\end{tabular}
\caption{Cross sections in the signal region for the
benchmark points P1 -- P8, as well as the corresponding
acceptances.}
\label{table:P:signal}
\end{center}
\end{table}

The $M_{bb}$ spectra are normalised to 100~fb$^{-1}$
integrated luminosity; since the event numbers per bin are small for
P2 and P4, several 100~fb$^{-1}$ integrated luminosity will be
required to see a statistically relevant number of signal events.
Then, the signals of a 125~GeV Higgs boson 
decaying into $b \bar{b}$ can be well visible above the
background, even for P2 and P4 despite the large
widths of the peak in this channel.

\begin{figure}[ht!]
\begin{center}
\hspace*{-6mm}
\begin{tabular}{cc}
\psfig{file=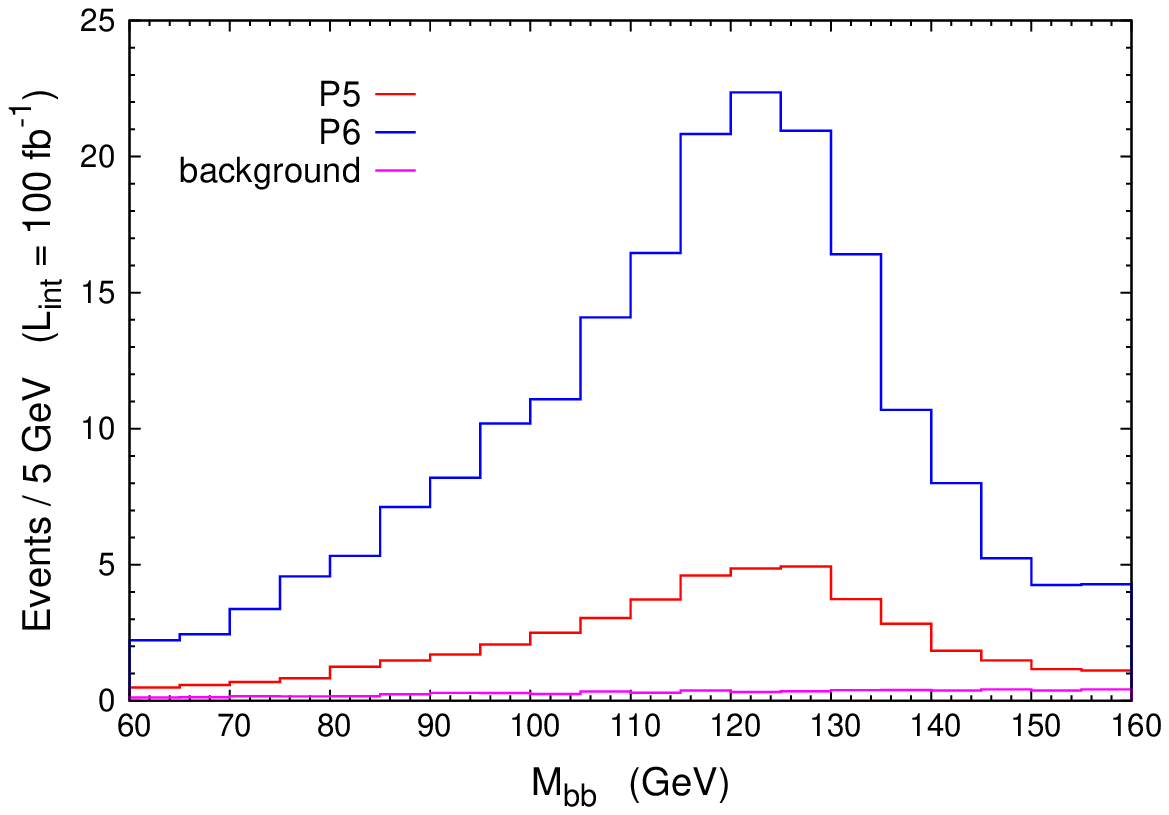, scale=0.60}
\   &
\psfig{file=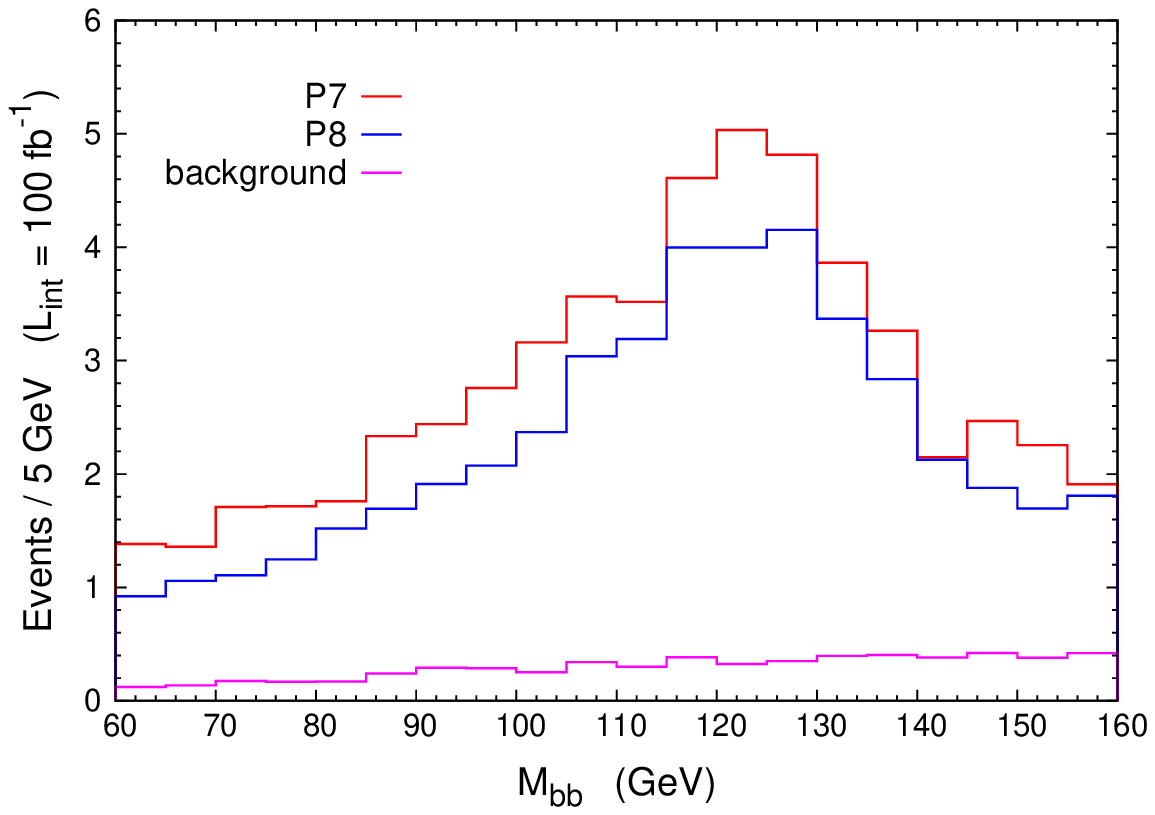, scale=0.60}
\end{tabular}
\end{center}
\caption{Spectra of $M_{bb}$ for the benchmark points P5 and P6
(left panel), and P7 and P8 (right panel).}
\label{fig:4}
\end{figure}

In Fig.~\ref{fig:4} we display the spectra of $M_{bb}$ after the cuts
described in Section~3 for the points P5 -- P8. The cross
sections in the signal region for these points are $\sim 0.45$~fb~(P5),
$\sim 2.0$~fb~(P6), $\sim 0.56$~fb~(P7) and $\sim 0.46$~fb~(P8), i.e.
still above the background cross section in the signal region
(see Table~\ref{table:P:signal}). Again, 
event numbers in the signal regions are small for P5, P7 and P8;
several 100~fb$^{-1}$ of integrated luminosity are required 
for a statistically relevant number of signal events. 
Given the acceptances of $\sim 1.4
\times 10^{-3}$ (for P8) to $\sim 3.3 \times 10^{-3}$ (for P7), the
signal rates -- if visible -- can provide at least rough information
on the initial squark/gluon cross sections, and thus on the strongly
interacting SUSY spectrum.

\begin{figure}[ht!]
\begin{center}
\hspace*{-6mm}
\begin{tabular}{cc}
\epsfig{file=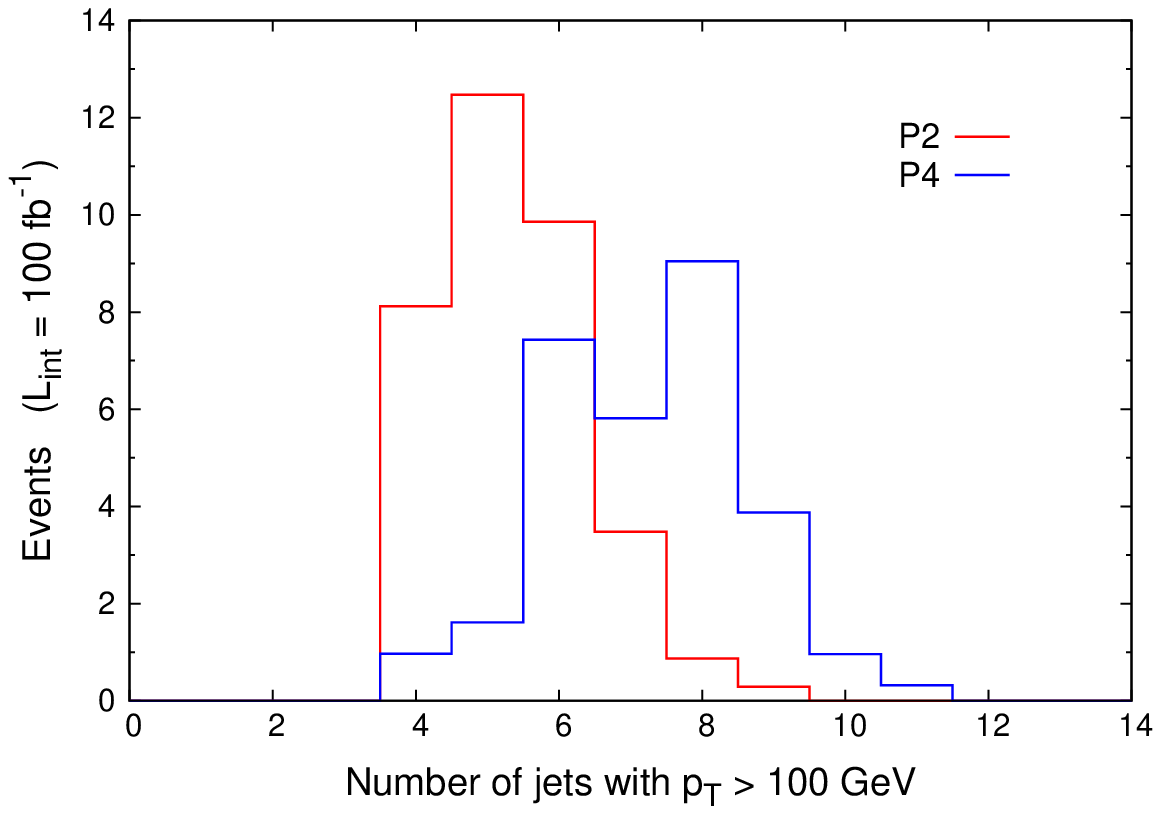, scale=0.60}
\   &
\epsfig{file=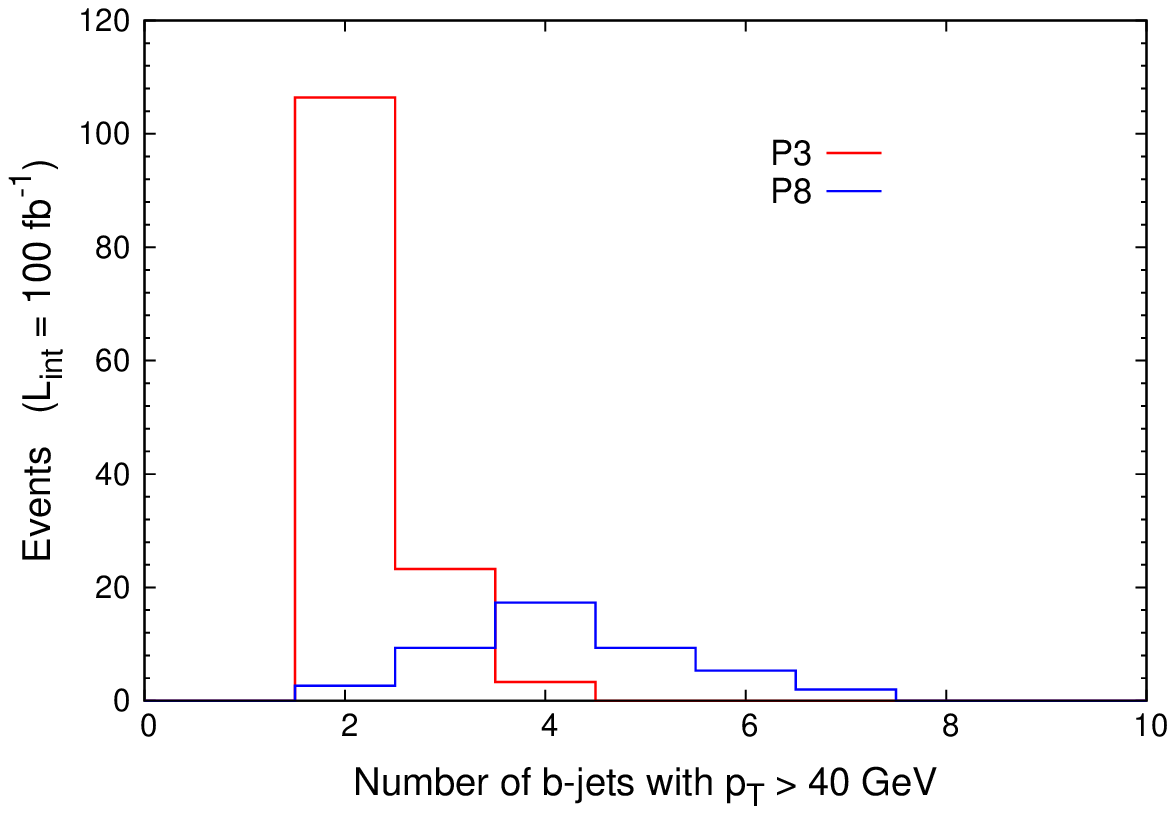, scale=0.60}
\end{tabular}
\end{center}
\caption{Hard jet (with $P_T > 100$~GeV) multiplicities for the benchmark
points P2 and P4 (left panel), and $b$-jet multiplicities for the
benchmark points P3 and P8 (right panel).}
\label{fig:5}
\end{figure}

Additional observables, which can help to shed some light on the nature of
the decay cascades, are the abundances of hard jets (with $P_T > 100$~GeV)
and $b$-jets (with $P_T > 40$~GeV) in the events which have passed the
cuts. Both observables differ considerably for the various benchmark points.
Extreme cases of high and low hard jet multiplicities are shown in the
left panel of Fig.~\ref{fig:5} for points P2 and P4; extreme cases
of high and low $b$-jet multiplicities are shown in the right panel of
Fig.~\ref{fig:5} for points P3 and P8 (all normalised to
100~fb$^{-1}$ integrated luminosity).

The jet multiplicities provide information on the decay cascades:
Large hard jet multiplicities appear generally for benchmark points with
gluinos lighter than squarks, such that squarks decay dominantly (or
exclusively) via gluinos. Instead of showing diagrams of these
multiplicities for all points, we find it convenient to define the ratio
$R_\mathrm{hard}$ of the number of events with 6 or more hard
jets (with $P_T > 100$~GeV) to the number of events with 5 or less
hard jets. This ratio is independent of the integrated luminosity
and suffers somewhat less from statistical fluctuations than the absolute
number of events per bin. (Of course, it still depends on the cuts
applied, and even on the used jet algorithm.)

The $b$-jet multiplicities indicate whether top or bottom squarks appear
in the gluino decay cascades, and are more pronounced if squarks decay
via gluinos. Including two $b$-jets from one Higgs boson 
(as the other Higgs necessarily decays into two $\tau$'s to satisfy
our cuts), each event for
P7 and P8 contains a priori six $b$-jets from the lowest order matrix 
element -- not all of which are tagged, but QCD radiation can
add more $b$~quark pairs. Again it is useful to define a ratio
$R_{b-\mathrm{jets}}$ of the number of events with 3 or more
$b$-jets (with $P_T > 40$~GeV) to the number of events with exactly 2 $b$-jets.
The ratios $R_\mathrm{hard}$ and $R_{b-\mathrm{jets}}$ are summarised
for the eight benchmark points in Table~3.

\begin{table}[ht!]
\begin{center}
\begin{tabular}{|c|c|c|c|c|c|c|c|c|}
\hline
 & P1 & P2 & P3 & P4 & P5 & P6 & P7 & P8 \\
\hline
$R_\mathrm{hard}$& 0.54 & 0.70 & 3.4 & 10.6 & 1.6 & 0.79 & 6.4 & 4.3 \\
\hline
$R_{b-\mathrm{jets}}$& 0.09 & 0.09 & 0.25 & 0.37 & 1.0 & 0.63 & 6.8 & 16.3 \\
\hline
\end{tabular}
\caption{The ratios $R_\mathrm{hard}$ and $R_{b-\mathrm{jets}}$ as defined
in the text for the eight benchmark points P1 -- P8.}
\end{center}
\end{table}

Despite the statistical fluctuations, the following trends can be
observed:
\begin{itemize}
\item Points P1, P2, P5 and P6, with gluinos heavier than squarks
(i.e. squarks decaying directly into a quark and the bino),
have $R_\mathrm{hard} \lsim 2$ (actually $\lsim 1$ except for
P5 with gluinos decaying into top/stop); $R_\mathrm{hard} \gsim 3$
indicates longer squark cascades via gluinos as for P3, P4, P7 and P8. Of
course, $R_\mathrm{hard}$ increases also with the squark masses, as is
visible when comparing P1/P2 and P3/P4.
\item Points P1, P2, P3 and P4 without stops/sbottoms in the gluino
decay cascades all have
$R_{b-\mathrm{jets}} \lsim 0.5$. Once gluinos can decay into 
stops/sbottoms, but for gluinos still heavier than squarks -- 
as for P5 and P6 --, we have
$0.5 \lsim R_{b-\mathrm{jets}} \lsim 1$. With stops/sbottoms in gluino
decays and squarks heavier than gluinos (as for P7 and P8), we have
$R_{b-\mathrm{jets}}\gsim 6$.
\end{itemize}
Finally, we recall that points P5/P7 with gluinos decaying via top/stop
have sizeable $E_T^\mathrm{miss}$ from leptonic top decays
(see Fig.~\ref{fig:2}), which allows to
distinguish them from points P6/P8, in which gluinos decay via
bottom/sbottom pairs.

These differences appear clear and easily distinguishable 
since the benchmark points
correspond to simplified models with (mostly) 100\% branching fractions
into given channels.
Still, together with the signal rates, the discussed observables would
give strong hints on the underlying sparticle spectrum.

\section{Summary and conclusions}
In the presence of a light singlino-like LSP in the NMSSM and an NLSP
with a mass slightly above the threshold for the NLSP~$\to$~LSP + Higgs
decay, the $E_T^\mathrm{miss}$ signature of sparticle production is
considerably reduced. In these scenarios the upper bounds on squark/gluino
masses from the run~I at the LHC are alleviated, and search strategies
not relying on large $E_T^\mathrm{miss}$ would also be required for the
run~II at 13~TeV c.m. energy.

We have proposed benchmark points corresponding to different squark/gluino
masses and decay cascades, and studied the prospects for search strategies
relying on two Higgs bosons in the final state. The proposed
cuts lead to large signal-to-background ratios for all masses and decay
cascades considered here, but can still be optimised
after realistic detector simulations. This may be required for squark/gluino
masses at or beyond the 1.4/1.5~TeV range, where the small signal cross sections
after our cuts would require several 100~fb$^{-1}$ of integrated luminosity;
looser cuts could be of help here.

The scenario with a light singlino-like LSP in the NMSSM would
also influence searches for direct stop/sbottom pair production, as well
as searches for direct neutralino/chargino production. In all these
cases, $E_T^\mathrm{miss}$ would be reduced as well, and two Higgs
bosons would appear instead in the decay cascades. Dedicated search strategies
for these cases for the run~II at the LHC remain to be devised.


\section*{Acknowledgements}

U.~E. and A.~M.~T. acknowledge support from European Union Initial
Training Network INVISIBLES (PITN-GA-2011-289442). U.~E.
acknowledges support from the ERC advanced grant Higgs@LHC, and from
the European Union Initial Training Network Higgs\-Tools
(PITN-GA-2012-316704).


\newpage

\begin{thebibliography}{99}

\bibitem{atlas_summary}
  G.~Aad {\it et al.}  [ATLAS Collaboration],
  JHEP {\bf 1310} (2013) 130
   [Erratum-ibid.\  {\bf 1401} (2014) 109]
  [arXiv:1308.1841 [hep-ex]], for an overview see
\nl {\sf 
https://twiki.cern.ch/twiki/bin/view/AtlasPublic/SupersymmetryPublicResults}

\bibitem{cms_summary} 
  S.~Chatrchyan {\it et al.}  [CMS Collaboration],
  JHEP {\bf 1406} (2014) 055
  [arXiv:1402.4770 [hep-ex]], for an overview see
\nl {\sf
https://twiki.cern.ch/twiki/bin/view/CMSPublic/PhysicsResultsSUS}

\bibitem{Feng:2013pwa}
  J.~L.~Feng,
  Ann.\ Rev.\ Nucl.\ Part.\ Sci.\  {\bf 63} (2013) 351
  [arXiv:1302.6587 [hep-ph]].

\bibitem{Craig:2013cxa}
  N.~Craig,
  ``The State of Supersymmetry after Run I of the LHC,''
  arXiv:1309.0528 [hep-ph].

\bibitem{Melzer-Pellmann:2014eta}
  I.~-A.~Melzer-Pellmann and P.~Pralavorio,
  ``Lessons for SUSY from the LHC after the first run,''
  arXiv:1404.7191 [hep-ex].

\bibitem{Halkiadakis:2014qda}
  E.~Halkiadakis, G.~Redlinger and D.~Shih,
  ``Status and Implications of BSM Searches at the LHC,''
  arXiv:1411.1427 [hep-ex].

\bibitem{Aad:2014wea}
  G.~Aad {\it et al.}  [ATLAS Collaboration],
  JHEP {\bf 1409} (2014) 176
  [arXiv:1405.7875 [hep-ex]].

\bibitem{Fan:2011yu}
  J.~Fan, M.~Reece and J.~T.~Ruderman,
  JHEP {\bf 1111} (2011) 012
  [arXiv:1105.5135 [hep-ph]].

\bibitem{Lisanti:2011tm}
  M.~Lisanti, P.~Schuster, M.~Strassler and N.~Toro, 
  JHEP {\bf 1211} (2012) 081
  [arXiv:1107.5055 [hep-ph]].

\bibitem{Fan:2012jf}
  J.~Fan, M.~Reece and J.~T.~Ruderman,
  JHEP {\bf 1207} (2012) 196
  [arXiv:1201.4875 [hep-ph]].
 	
\bibitem{Baryakhtar:2012rz}
  M.~Baryakhtar, N.~Craig and K.~Van Tilburg,
  JHEP {\bf 1207} (2012) 164
  [arXiv:1206.0751 [hep-ph]].

\bibitem{Evans:2013jna}
  J.~A.~Evans, Y.~Kats, D.~Shih and M.~J.~Strassler,
  JHEP {\bf 1407} (2014) 101
  [arXiv:1310.5758 [hep-ph]].

\bibitem{Ellwanger:2014hia}
  U.~Ellwanger and A.~M.~Teixeira,
  JHEP {\bf 1410} (2014) 113
  [arXiv:1406.7221 [hep-ph]].

\bibitem{Ellwanger:2009dp}
  U.~Ellwanger, C.~Hugonie and A.~M.~Teixeira,
  Phys.\ Rept.\  {\bf 496} (2010) 1\newline
  [arXiv:0910.1785 [hep-ph]].

\bibitem{Kim:1983dt}
  J.~E.~Kim and H.~P.~Nilles,
  Phys.\ Lett.\ B {\bf 138} (1984) 150.

\bibitem{Hall:2011aa}
  L.~J.~Hall, D.~Pinner and J.~T.~Ruderman,
  JHEP {\bf 1204} (2012) 131
  [arXiv:1112.2703 [hep-ph]].
  
\bibitem{Ellwanger:2011aa}
  U.~Ellwanger,
  JHEP {\bf 1203} (2012) 044
  [arXiv:1112.3548 [hep-ph]].
  
\bibitem{Arvanitaki:2011ck}
  A.~Arvanitaki and G.~Villadoro,
  JHEP {\bf 1202} (2012) 144
  [arXiv:1112.4835 [hep-ph]].
  
\bibitem{King:2012is}
  S.~F.~King, M.~Muhlleitner and R.~Nevzorov,
  Nucl.\ Phys.\ B {\bf 860} (2012) 207
  [arXiv:1201.2671 [hep-ph]].

\bibitem{Kang:2012sy}
  Z.~Kang, J.~Li and T.~Li,
  JHEP {\bf 1211} (2012) 024
  [arXiv:1201.5305 [hep-ph]].
  
\bibitem{Cao:2012fz}
  J.~-J.~Cao, Z.~-X.~Heng, J.~M.~Yang, Y.~-M.~Zhang and J.~-Y.~Zhu,
  JHEP {\bf 1203} (2012) 086
  [arXiv:1202.5821 [hep-ph]].

\bibitem{ATLAS-CONF-2013-091}
  The ATLAS collaboration,
  ``Search for massive particles decaying into multible quarks with the
   ATLAS Detector in $\sqrt{s}=8$~TeV $pp$ collisions,''
  ATLAS-CONF-2013-091.

\bibitem{CMS-PAS-EXO-12-049} CMS collaboration,
 ``Search for light- and heavy-flavour three-jet resonances in multijet
    final states at 8~TeV,'' CMS-PAS-EXO-12-049.

\bibitem{Chatrchyan:2013izb}
  S.~Chatrchyan {\it et al.}  [CMS Collaboration],
  Phys.\ Rev.\ Lett.\  {\bf 110} (2013) 141802
  [arXiv:1302.0531 [hep-ex]].

\bibitem{Chatrchyan:2013xva}
  S.~Chatrchyan {\it et al.}  [CMS Collaboration],
  JHEP {\bf 1307} (2013) 178
  [arXiv:1303.5338 [hep-ex]].

\bibitem{Dimopoulos:1996vz}
  S.~Dimopoulos, M.~Dine, S.~Raby and S.~D.~Thomas,
  Phys.\ Rev.\ Lett.\  {\bf 76} (1996) 3494
  [hep-ph/9601367].

\bibitem{Ambrosanio:1996jn}
  S.~Ambrosanio, G.~L.~Kane, G.~D.~Kribs, S.~P.~Martin and S.~Mrenna,
  Phys.\ Rev.\ D {\bf 54} (1996) 5395
  [hep-ph/9605398]. 

\bibitem{Hinchliffe:1996iu}
  I.~Hinchliffe, F.~E.~Paige, M.~D.~Shapiro, J.~Soderqvist and W.~Yao,
  Phys.\ Rev.\ D {\bf 55} (1997) 5520
  [hep-ph/9610544].

\bibitem{Matchev:1999ft}
  K.~T.~Matchev and S.~D.~Thomas,
  Phys.\ Rev.\ D {\bf 62} (2000) 077702
  [hep-ph/9908482].

\bibitem{Datta:2003iz}
  A.~Datta, A.~Djouadi, M.~Guchait and F.~Moortgat,
  Nucl.\ Phys.\ B {\bf 681} (2004) 31
  [hep-ph/0303095].

\bibitem{Bandyopadhyay:2008fp}
  P.~Bandyopadhyay, A.~Datta and B.~Mukhopadhyaya,
  Phys.\ Lett.\ B {\bf 670} (2008) 5
  [arXiv:0806.2367 [hep-ph]].

\bibitem{Huitu:2008sa}
  K.~Huitu, R.~Kinnunen, J.~Laamanen, S.~Lehti, S.~Roy and T.~Salminen,
  Eur.\ Phys.\ J.\ C {\bf 58} (2008) 591
  [arXiv:0808.3094 [hep-ph]].

\bibitem{Bandyopadhyay:2008sd}
  P.~Bandyopadhyay,
  JHEP {\bf 0907} (2009) 102
  [arXiv:0811.2537 [hep-ph]].

\bibitem{Fowler:2009ay}
  A.~C.~Fowler and G.~Weiglein,
  JHEP {\bf 1001} (2010) 108
  [arXiv:0909.5165 [hep-ph]].
 	
\bibitem{Meade:2009qv}
  P.~Meade, M.~Reece and D.~Shih,
  JHEP {\bf 1005} (2010) 105
  [arXiv:0911.4130 [hep-ph]].

\bibitem{Asano:2010ut}
  M.~Asano, H.~D.~Kim, R.~Kitano and Y.~Shimizu,
  JHEP {\bf 1012} (2010) 019\nl
  [arXiv:1010.0692 [hep-ph]].
 	
\bibitem{Thaler:2011me}
  J.~Thaler and Z.~Thomas,
  JHEP {\bf 1107} (2011) 060
  [arXiv:1103.1631 [hep-ph]].

\bibitem{Gori:2011hj}
  S.~Gori, P.~Schwaller and C.~E.~M.~Wagner,
  Phys.\ Rev.\ D {\bf 83} (2011) 115022
  [arXiv:1103.4138 [hep-ph]].
 	
\bibitem{Ruderman:2011vv}
  J.~T.~Ruderman and D.~Shih,
  JHEP {\bf 1208} (2012) 159
  [arXiv:1103.6083 [hep-ph]].

\bibitem{Kats:2011qh}
  Y.~Kats, P.~Meade, M.~Reece and D.~Shih,
  JHEP {\bf 1202} (2012) 115
  [arXiv:1110.6444 [hep-ph]].

\bibitem{Baer:2012ts}
  H.~Baer, V.~Barger, A.~Lessa, W.~Sreethawong and X.~Tata,
  Phys.\ Rev.\ D {\bf 85} (2012) 055022
  [arXiv:1201.2949 [hep-ph]].

\bibitem{Ghosh:2012mc}
  D.~Ghosh, M.~Guchait and D.~Sengupta,
  Eur.\ Phys.\ J.\ C {\bf 72} (2012) 2141 
  [arXiv:1202.4937 [hep-ph]].

\bibitem{Belyaev:2012si}
  A.~Belyaev, J.~P.~Hall, S.~F.~King and P.~Svantesson,
  Phys.\ Rev.\ D {\bf 86} (2012) 031702
  [arXiv:1203.2495 [hep-ph]].

\bibitem{Byakti:2012qk}
  P.~Byakti and D.~Ghosh,
  Phys.\ Rev.\ D {\bf 86} (2012) 095027
  [arXiv:1204.0415 [hep-ph]].

\bibitem{Howe:2012xe}
  K.~Howe and P.~Saraswat,
  JHEP {\bf 1210} (2012) 065
  [arXiv:1208.1542 [hep-ph]].

\bibitem{Belyaev:2012jz}
  A.~Belyaev, J.~P.~Hall, S.~F.~King and P.~Svantesson,
  Phys.\ Rev.\ D {\bf 87} (2013) 3,  035019
  [arXiv:1211.1962 [hep-ph]].
	
\bibitem{Arbey:2012fa}
  A.~Arbey, M.~Battaglia and F.~Mahmoudi,
  ``Higgs Production in Neutralino Decays in the MSSM - The LHC and a
  Future e+e- Collider,'' 
  arXiv:1212.6865 [hep-ph].

\bibitem{Bharucha:2013epa}
  A.~Bharucha, S.~Heinemeyer and F.~von der Pahlen,
  Eur.\ Phys.\ J.\ C {\bf 73} (2013) 2629
  [arXiv:1307.4237].

\bibitem{Han:2013kza}
  T.~Han, S.~Padhi and S.~Su,
  Phys.\ Rev.\ D {\bf 88} (2013) 115010
  [arXiv:1309.5966 [hep-ph]].

\bibitem{Yu:2014mda}
  F.~Yu,
  Phys.\ Rev.\ D {\bf 90} (2014) 015009
  [arXiv:1404.2924 [hep-ph]].

\bibitem{Franke:1995tf}
  F.~Franke and H.~Fraas,
  Z.\ Phys.\ C {\bf 72} (1996) 309
  [hep-ph/9511275].

\bibitem{Ellwanger:1997jj}
  U.~Ellwanger and C.~Hugonie,
  Eur.\ Phys.\ J.\ C {\bf 5} (1998) 723
  [hep-ph/9712300].

\bibitem{Choi:2004zx}
  S.~Y.~Choi, D.~J.~Miller and P.~M.~Zerwas,
  Nucl.\ Phys.\ B {\bf 711} (2005) 83
  [hep-ph/0407209].

\bibitem{Cheung:2008rh}
  K.~Cheung and T.~J.~Hou,
  Phys.\ Lett.\ B {\bf 674} (2009) 54
  [arXiv:0809.1122 [hep-ph]].
 	
\bibitem{Stal:2011cz}
  O.~Stal and G.~Weiglein,
  JHEP {\bf 1201} (2012) 071
  [arXiv:1108.0595 [hep-ph]].

\bibitem{Das:2012rr}
  D.~Das, U.~Ellwanger and A.~M.~Teixeira,
  JHEP {\bf 1204} (2012) 067
  [arXiv:1202.5244 [hep-ph]].

\bibitem{Cerdeno:2013qta}
  D.~G.~Cerde\~no, P.~Ghosh, C.~B.~Park and M.~Peiró,
  JHEP {\bf 1402} (2014) 048
  [arXiv:1307.7601 [hep-ph], arXiv:1307.7601].

\bibitem{Kribs:2009yh}
  G.~D.~Kribs, A.~Martin, T.~S.~Roy and M.~Spannowsky,
  Phys.\ Rev.\ D {\bf 81} (2010) 111501
  [arXiv:0912.4731 [hep-ph]].

\bibitem{Kribs:2010hp}
  G.~D.~Kribs, A.~Martin, T.~S.~Roy and M.~Spannowsky,
  Phys.\ Rev.\ D {\bf 82} (2010) 095012
  [arXiv:1006.1656 [hep-ph]].

\bibitem{Bhattacherjee:2012bu}
  B.~Bhattacherjee, A.~Chakraborty, D.~Kumar Ghosh and S.~Raychaudhuri,
  Phys.\ Rev.\ D {\bf 86} (2012) 075012
  [arXiv:1204.3369 [hep-ph]].

\bibitem{Ellwanger:2004xm}
  U.~Ellwanger, J.~F.~Gunion and C.~Hugonie,
  JHEP {\bf 0502} (2005) 066
  [hep-ph/0406215].

\bibitem{Ellwanger:2005dv}
  U.~Ellwanger and C.~Hugonie,
  Comput.\ Phys.\ Commun.\  {\bf 175} (2006) 290
  [hep-ph/0508022].

\bibitem{Drees:2013wra}
  M.~Drees, H.~Dreiner, D.~Schmeier, J.~Tattersall and J.~S.~Kim,
  ``CheckMATE: Confronting your Favourite New Physics Model with LHC Data,''
  arXiv:1312.2591 [hep-ph].

\bibitem{Beenakker:1996ch}
  W.~Beenakker, R.~Hopker, M.~Spira and P.~M.~Zerwas,
  Nucl.\ Phys.\ B {\bf 492} (1997) 51
  [hep-ph/9610490].

\bibitem{Beenakker:1996ed}
  W.~Beenakker, R.~Hopker and M.~Spira,
  ``PROSPINO: A Program for the production of supersymmetric particles
  in next-to-leading order QCD,'' 
  hep-ph/9611232.

\bibitem{Alwall:2011uj}
  J.~Alwall, M.~Herquet, F.~Maltoni, O.~Mattelaer and T.~Stelzer,
  JHEP {\bf 1106} (2011) 128
  [arXiv:1106.0522 [hep-ph]].

\bibitem{Sjostrand:2006za}
  T.~Sjostrand, S.~Mrenna and P.~Z.~Skands,
  JHEP {\bf 0605} (2006) 026
  [hep-ph/0603175].

\bibitem{deFavereau:2013fsa}
  J.~de Favereau {\it et al.}  [DELPHES 3 Collaboration],
  JHEP {\bf 1402} (2014) 057\nl
  [arXiv:1307.6346 [hep-ex]].

\bibitem{Cacciari:2011ma}
  M.~Cacciari, G.~P.~Salam and G.~Soyez,
  Eur.\ Phys.\ J.\ C {\bf 72} (2012) 1896
  [arXiv:1111.6097 [hep-ph]].

\bibitem{Cacciari:2008gp}
  M.~Cacciari, G.~P.~Salam and G.~Soyez,
  JHEP {\bf 0804} (2008) 063
  [arXiv:0802.1189 [hep-ph]].

\bibitem{Conte:2012fm}
  E.~Conte, B.~Fuks and G.~Serret,
  Comput.\ Phys.\ Commun.\  {\bf 184} (2013) 222\nl
  [arXiv:1206.1599 [hep-ph]].

\bibitem{Conte:2013mea}
  E.~Conte and B.~Fuks,
  ``MadAnalysis 5: status and new developments,''\nl
  arXiv:1309.7831 [hep-ph].












\end{thebibliography}
\end{document}